\documentclass[manuscript]{aastex}
\usepackage{epsfig}
\usepackage{graphicx}% Include figure files
\usepackage{dcolumn}% Align table columns on decimal point
\usepackage{bm}% bold math
\usepackage{amsmath}
\usepackage{cite}
\usepackage[
top    = 2.75cm,
bottom = 2.50cm,
left   = 1.50cm,
right  = 1.50cm]
{geometry}

\shorttitle{Non-extensive Statistics Solution to the Cosmological Lithium Problem}
\shortauthors{S.Q. Hou et al.}

\usepackage{xcolor}
\begin{document}


\title{Non-extensive Statistics Solution to the Cosmological Lithium Problem}


\author{S.Q. Hou\altaffilmark{1}, J.J. He\altaffilmark{1,2}, A. Parikh\altaffilmark{3,4}, D. Kahl\altaffilmark{5,6}\\ C.A. Bertulani\altaffilmark{7},
T. Kajino\altaffilmark{8,9,10}, G.J. Mathews\altaffilmark{9,11}, G. Zhao\altaffilmark{2}}
\affil{\altaffilmark{1}Key Laboratory of High Precision Nuclear Spectroscopy, Institute of Modern Physics, Chinese Academy of Sciences, Lanzhou 730000, China}
\affil{\altaffilmark{2}Key Laboratory of Optical Astronomy, National Astronomical Observatories, Chinese Academy of Sciences, Beijing 100012, China}
\affil{\altaffilmark{3}Departament de F\'{\i}sica i Enginyeria Nuclear, EUETIB, Universitat Polit\`{e}cnica de Catalunya, Barcelona E-08036, Spain}
\affil{\altaffilmark{4}Institut d'Estudis Espacials de Catalunya, Barcelona E-08034, Spain}
\affil{\altaffilmark{5}Center for Nuclear Study, The University of Tokyo, RIKEN campus, Wako, Saitama 351-0198, Japan}
\affil{\altaffilmark{6}School of Physics \& Astronomy, the University of Edinburgh, Edinburgh EH9 3JZ, UK}
\affil{\altaffilmark{7}Texas A\&M University-Commerce, Commerce, TX 75429-3011, USA}
\affil{\altaffilmark{8}Department of Astronomy, School of Science, the University of Tokyo, 7-3-1 Hongo, Bunkyo-ku, Tokyo, 113-0033, Japan}
\affil{\altaffilmark{9}National Astronomical Observatory of Japan 2-21-1 Osawa, Mitaka, Tokyo, 181-8588, Japan}
\affil{\altaffilmark{10}International Research Center for Big-Bang Cosmology and Element Genesis, School of Physics and Nuclear Energy Engineering, Beihang University, Beijing 100191, China}
\affil{\altaffilmark{11}Center for Astrophysics, Department of Physics, University of Notre Dame, Notre Dame, IN 46556, USA}

\email{Corresponding author email: hejianjun@nao.cas.cn}

\begin{abstract}
Big Bang nucleosynthesis (BBN) theory predicts the abundances of the light elements D, $^3$He, $^4$He and $^7$Li produced in the early universe. The primordial
abundances of D and $^4$He inferred from observational data are in good agreement with predictions, however, the BBN theory overestimates the primordial $^7$Li
abundance by about a factor of three. This is the so-called ``cosmological lithium problem''. Solutions to this problem using conventional astrophysics and nuclear
physics have not been successful over the past few decades, probably indicating the presence of new physics during the era of BBN. We have investigated the impact on
BBN predictions of adopting a generalized distribution to describe the velocities of nucleons in the framework of Tsallis non-extensive statistics. This generalized
velocity distribution is characterized by a parameter $q$, and reduces to the usually assumed Maxwell-Boltzmann distribution for $q$ = 1. We find excellent agreement
between predicted and observed primordial abundances of D, $^4$He and $^7$Li for $1.069\leq q \leq 1.082$, suggesting a possible new solution to the cosmological
lithium problem.
\end{abstract}

\keywords{cosmology: early universe --- cosmology: primordial nucleosynthesis --- plasmas}

\section{Introduction}
First proposed in 1946 by George Gamow~\citep{gam46}, the hot Big-Bang theory is now the most widely accepted cosmological model of the universe, where the universe
expanded from a very high density state dominated by radiation. The theory has been vindicated by the observation of the cosmic microwave background~\citep{pen65,WMAP9},
our emerging knowledge on the large-scale structure of the universe, and the rough consistency between calculations and observations of primordial abundances of the
lightest elements in nature: hydrogen, helium, and lithium. The primordial Big-Bang Nucleosynthesis (BBN) began when the universe was 3-minutes old and ended less than
half an hour later when nuclear reactions were quenched by the low temperature and density conditions in the expanding universe. Only the lightest nuclides ($^2$H,
$^3$He, $^4$He, and $^7$Li) were synthesized in appreciable quantities through BBN, and these relics provide us a unique window on the early universe. The primordial
abundances of $^2$H (referred to as D hereafter) and $^4$He inferred from observational data are in good general agreement with predictions; however, the BBN theory
overestimates the primordial $^7$Li abundance by about a factor of three~\citep{cyb03,coc04,asp06,sbo10}. This is the so-called ``cosmological lithium problem''.
Attempts to resolve this discrepancy using conventional nuclear physics have been unsuccessful over the past few decades~\citep{ang05,cyb08,boy10,wan11,sch11,kir11,tor12,coc12,ham13,piz14,fam16},
although the nuclear physics solutions altering the reaction flow into and out of
mass-7 are still being proposed~\citep{cyb09,cha11}. The dire potential impact of this longstanding issue on our understanding of the early universe has prompted the
introduction of various exotic scenarios involving, for example, the introduction of new particles and interactions beyond the Standard
Model~\citep{pos10,kan12,coc13,yam14,kus14,and16}. On the observational side, there are attempts to improve our understanding of lithium depletion mechanisms operative
in stellar models~\citep{vau98,pin99,pin02,ric05,kor06}. This remains an important goal but is not our focus here. For the recent reviews on BBN and primordial lithium
problem, please read articles written by~\citet{fie11} and~\citet{cyb16}.

In this work we suggest one solution to the lithium problem that arises in a straightforward, simple manner from a modification of the velocity distributions of
nuclei during the era of BBN. In the BBN model, the predominant nuclear-physics inputs are thermonuclear reaction rates (derived from cross sections). In the past
decades, great efforts have been undertaken to determine these data with high accuracy (e.g., see compilations of~\citet{wag69,cf88,smi93,ang99,des04,ser04,xu13}). A key
assumption in all thermonuclear rate determinations is that the velocities of nuclei may be described by the classical Maxwell-Boltzmann (MB)
distribution~\citep{rol88,ili07}. The MB distribution was derived for describing the thermodynamic equilibrium properties of the ideal gas, and was verified by a
high-resolution experiment at a temperature of $\sim$~900 K about 60 years ago~\citep{mil55}. However, it is worth asking: Do nuclei still obey the classical MB
distribution in the extremely complex, fast-expanding, Big-Bang hot plasma? Indeed, \citet{cly75} adopted a similar approach when addressing the
solar neutrino problem prior to the unambiguous measurement of neutrino flavor change by~\citet{ahm01}.

Whatever the source of the distortions from MB, one expects that the distribution should still maximize entropy. Hence, to account for modifications to the classical
MB velocity distribution, one may use Tsallis statistics (also referred to as non-extensive statistics)~\citep{tsa88}, which is based on the concept of generalized
non-extensive entropy. The associated generalized velocity distribution is characterized by a parameter $q$ and reduces to the MB distribution for $q = 1$. Tsallis
statistics has been applied in a host of different fields, including physics, astronomy, biology and economics~\citep{gel04}.

\section{Thermonuclear reaction rate}
It is well-known that thermonuclear rate for a typical $1+2\rightarrow3+4$ reaction is usually calculated by folding the cross section $\sigma(E)_{12}$ with a MB
distribution~\citep{rol88,ili07}
%\begin{widetext}
\begin{equation}
\label{eq1}
\left\langle\sigma v\right\rangle_{12}=\sqrt{\frac{8}{\pi\mu_{12}(kT)^3}}\int_{0}^{\infty}\sigma(E)_{12}E\mathrm{exp}\left(-\frac{E}{kT}\right)\,dE,
\end{equation}
%\end{widetext}
with $k$ the Boltzmann constant, $\mu_{12}$ the reduced mass of particles $1$ and $2$.
In Tsallis statistics, the velocity distribution of particles can be expressed by~\citet{tsa88}
\begin{equation}
\label{eq2}
f_q(\mathbf{v})=B_q\left(\frac{m}{2\pi kT}\right)^{3/2}\left[1-(q-1)\frac{m\mathbf{v}^2}{2kT}\right]^{\frac{1}{q-1}},
\end{equation}
where $B_q$ denotes the $q$-dependent normalization constant. With this velocity distribution, the non-extensive thermonuclear rate~\citep{ili07} for a typical
$1+2\rightarrow3+4$ reaction, where both reactants and products are nuclei, can be calculated by:
%\begin{widetext}
\begin{equation}
\label{eq3}
\left\langle\sigma v\right\rangle_{12}= B_q\sqrt{\frac{8}{\pi\mu_{12}}}\times\frac{1}{(kT)^{3/2}}\times\int_{0}^{E_\mathrm{max}}\sigma_{12}(E)E\left[1-(q-1)\frac{E}{kT}\right]^{\frac{1}{q-1}}\,dE,
\end{equation}
%\end{widetext}
with $E_\mathrm{max}$=$\frac{kT}{q-1}$ for $q>1$ and +$\infty$ for $0<q<1$. Here, the $q<0$ case is excluded according to the maximum-entropy
principle~\citep{tsa88,gel04}. Usually, one defines the $1+2\rightarrow3+4$ reaction with positive $Q$ value as the forward reaction and the corresponding
$3+4\rightarrow1+2$ reaction with negative $Q$ value as the reverse one. Under the assumption of classical statistics, the ratio between reverse and forward rates is
simply proportional to exp$(-\frac{Q}{kT})$~\citep{ili07}. With Tsallis statistics, however, the reverse rate is expressed as:
%\begin{widetext}
\begin{equation}
%\begin{eqnarray*}
\label{eq4}
\left\langle\sigma v\right\rangle_{34}=c\times B_q\sqrt{\frac{8}{\pi\mu_{12}}}\times\frac{1}{(kT)^{3/2}}\times\int_{0}^{E_\mathrm{max}-Q}\sigma_{12}(E)E\left[1-(q-1)\frac{E+Q}{kT}\right]^{\frac{1}{q-1}}\,dE,
%\end{eqnarray*}
\end{equation}
%\end{widetext}
where $c$=$\frac{(2J_1+1)(2J_2+1)(1+\delta_{34})}{(2J_3+1)(2J_4+1)(1+\delta_{12})}(\frac{\mu_{12}}{\mu_{34}})^{3/2}$. All parameters in Eqs.~(1--3) are well-defined
in~\citet{ili07}. For a reaction $1+2 \rightarrow 3+\gamma$, we assume the photons obey the Planck radiation law~\citep{ili07,tor97,tor98} and use the approximation
of $e^{E\gamma/kT}-1\approx e^{E\gamma/kT}$~\citep{mat11} when calculating the corresponding reverse rate.

\begin{figure}[t]
\begin{center}
\includegraphics[width=8.6cm]{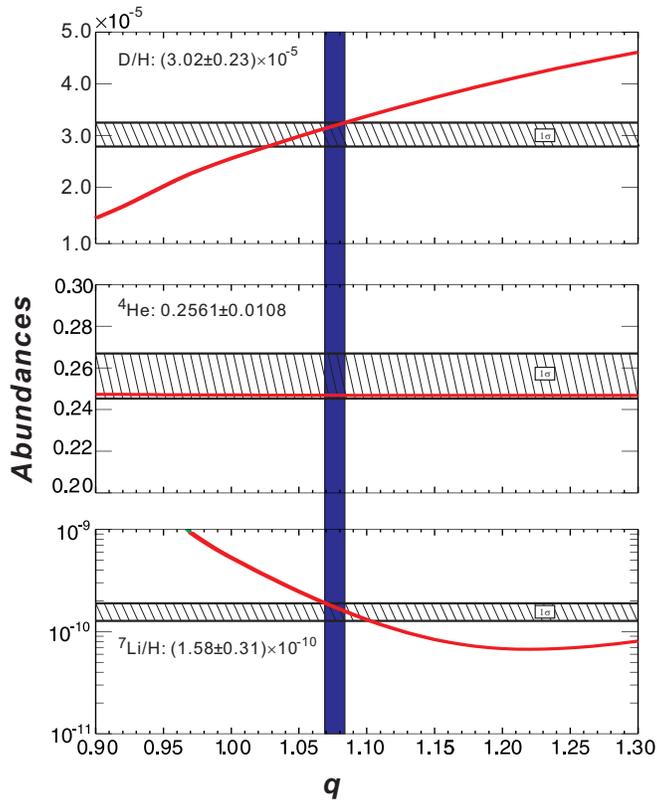}
\vspace{-1mm} \caption{\label{fig1} Predicted primordial abundances as a function of parameter $q$ (in red solid lines). The observed primordial
abundances~\citep{oli12,ave10,sbo10} with 1$\sigma$ uncertainty for D, $^4$He, and $^7$Li are indicated as hatched horizontal bands. The vertical (blue) band constrains
the range of the parameter $q$ to $1.069\leq q \leq 1.082$. Note that the `abundance' of $^4$He exactly refers to its mass fraction.}
\end{center}
\end{figure}

\renewcommand{\arraystretch}{0.85}
\begin{table*}
\scriptsize
%\begin{center}
%\setlength{\belowcaptionskip}{10pt}
\caption{\label{tab1} Nuclear reactions involved in the present BBN network. The non-extensive Tsallis distribution is implemented for the 17 principal reactions shown
in the bold face. The listed flux Ratio is the time-integrated reaction flux calculated with the non-extensive Tsallis distribution (with $q$ = 1.0755) relative to
that with the classical MB distribution ($q$ = 1). The references are listed for each reaction in the square brackets.}
%\begin{ruledtabular}
\begin{tabular}{lr|lr|}
%\begin{tabular*}{\linewidth}{@{\hspace{2mm}\extracolsep{\fill}}lr|lr@{\hspace{2mm}}}
\hline
\hline
Reaction & Ratio & Reaction & Ratio \\
\hline
\textbf{$^1$H($n$,$\gamma$)$^2$H}~\citep{har03} 	                & 1.02 & $^2$H($n$,$\gamma$)$^3$H~\citep{wag69}	      & 1.09  \\
\textbf{$^2$H($p$,$\gamma$)$^3$He}~\citep{des04}	                & 0.81 & $^3$He($n$,$\gamma$)$^4$He~\citep{wag69}	  & 1.10  \\
\textbf{$^2$H($d$,$n$)$^3$He}~\citep{des04}	                    & 1.12 & $^3$He($^3$He,2$p$)$^4$He~\citep{cf88}     	  & 1.54  \\
\textbf{$^2$H($d$,$p$)$^3$H}~\citep{des04}	                    & 0.91 & 2$^4$He($n$,$\gamma$)$^9$Be~\citep{cf88}      & 0.62  \\
\textbf{$^3$H($d$,$n$)$^4$He}~\citep{des04}	                    & 1.02 & $^6$Li($p$,$\gamma$)$^7$Be~\citep{xu13,hjj13} & 0.59  \\
\textbf{$^3$H($\alpha$,$\gamma$)$^7$Li}~\citep{des04}            & 0.60 & $^6$Li($n$,$\gamma$)$^7$Li~\citep{mal89}      & 0.47  \\
\textbf{$^3$He($n$,$p$)$^3$H}~\citep{des04}                      & 1.11 & $^6$Li($n$,$\alpha$)$^3$H~\citep{cf88}        & 0.47  \\
\textbf{$^3$He($d$,$p$)$^4$He}~\citep{des04}                     & 0.84 & $^7$Li($n$,$\gamma$)$^8$Li~\citep{wag69}      & 1.06  \\
\textbf{$^3$He($\alpha$,$\gamma$)$^7$Be}~\citep{des04}           & 0.37 & $^8$Li($n$,$\gamma$)$^9$Li~\citep{li05}       & 1.06  \\
\textbf{$^7$Li($p$,$\alpha$)$^4$He}~\citep{des04}                & 0.61 & $^8$Li($p$,$n$)2$^4$He~\citep{wag69}          & 1.07  \\
\textbf{$^7$Be($n$,$p$)$^7$Li }~\citep{des04}                    & 0.39 & $^9$Li($p$,$\alpha$)$^6$He~\citep{tho93}      & 1.07  \\
\textbf{$^3$H($p$,$\gamma$)$^4$He}~\citep{dub09}                 & 0.69 & $^9$Be($p$,$\alpha$)$^6$Li~\citep{cf88}       & 1.01  \\
\textbf{$^2$H($\alpha$,$\gamma$)$^6$Li}~\citep{ang99,xu13,and14} & 0.43 & $^9$Be($p$,$d$)2$^4$He~\citep{cf88}           & 0.97  \\
\textbf{$^6$Li($p$,$\alpha$)$^3$He}~\citep{ang99,xu13}           & 0.36 &                                              &       \\
\textbf{$^7$Be($n$,$\alpha$)$^4$He}~\citep{kin77}                & 0.35 &                                              &       \\
\textbf{$^7$Li($d$,$n$)2$^4$He}~\citep{cf88}                     & 0.53 &                                              &       \\
\textbf{$^7$Be($d$,$p$)2$^4$He}~\citep{cf88,par72}               & 0.11 &                                              &       \\
\hline \hline
\end{tabular}
%\end{ruledtabular}
\end{table*}
\renewcommand{\arraystretch}{1.0}

\section{Impact of non-extensive statistics on BBN}
A previous attempt to examine the impact of deviations from the MB distribution on BBN~\citep{ber13} only used non-extensive statistics for forward rates and did not
consider the impact on reverse rates. Here, we have for the first time used a non-extensive velocity distribution to determine thermonuclear reaction rates of primary
importance to BBN in a consistent manner. With these non-extensive rates, the primordial abundances are predicted by a standard BBN code by adopting the up-to-date
cosmological parameter $\eta$ = (6.203$\pm$0.137)$\times$10$^{-10}$~\citep{WMAP9} for the baryon-to-photon ratio, and the neutron lifetime of
$\tau_n$ = (880.3$\pm$1.1) s~\citep{oli14}. The reaction network involves 30 reactions in total with nuclei of $A \leq 9$  (see Table~\ref{tab1}). Here, the
thermonuclear (forward and reverse) rates for those 17 principal reactions (with bold face in Table~\ref{tab1}) have been determined in the present work using
non-extensive statistics, with 11 reactions of primary importance~\citep{smi93} and 6 of secondary importance~\citep{ser04} in the primordial light-element
nucleosynthesis. The standard MB rates have been adopted for the remaining reactions, as they play only a minor role during BBN. Our code gives results in good
agreement with the previous BBN predictions~\citep{ber13,coc12,cyb16} if $q$ = 1, as seen in Table~\ref{tab2}.

\begin{figure}[tbp]
\begin{center}
\includegraphics[width=8.6cm]{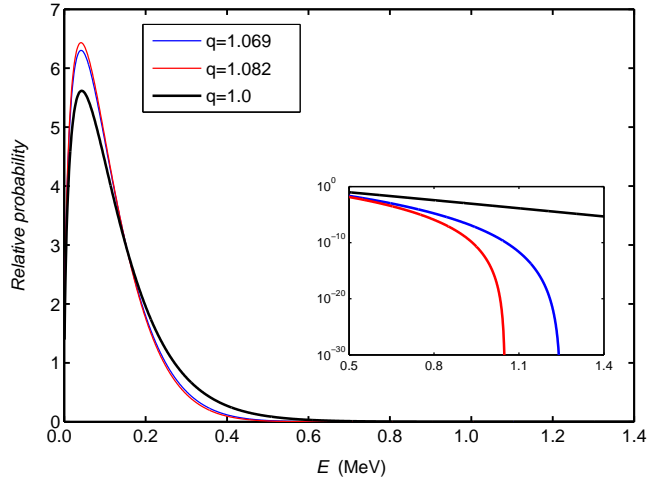}
\vspace{-1mm}
\caption{\label{fig2} Normalized relative probabilities for non-extensive energy distributions and for the standard MB distribution ($q$ = 1) at temperature of 1 GK.
The enlarged insert plot shows the tails, which are cut off at $E_\mathrm{max}=kT/(q-1)$ for the non-extensive distributions.}
\end{center}
\end{figure}

It shows that the predicted and observed abundances~\citep{oli12,ave10,sbo10} of D, $^4$He and $^7$Li fall into agreement (within 1$\sigma$ uncertainty of observed
data) when a non-extensive velocity distribution with $1.069\leq q \leq 1.082$ is adopted, as shown in Fig.~\ref{fig1} and Table~\ref{tab2}. As the reliability of
primordial $^3$He observations is still under debate~\citep{coc12}, we do not include this species in the figure. In this calculation, the predicted $^3$He abundance
for the above range of $q$ agrees at the 1.8$\sigma$ level with an abundance of $^3$He/H = 1.1(2)~\citep{ban02} observed in our Galaxy's interstellar medium.
Thus, we have found a possible new solution to the cosmological lithium problem without introducing any exotic theory. Figure 2 illustrates the level of deviation
from the MB energy distribution implied by $q$ = 1.069 and 1.082 at 1 GK.

\begin{table*}
\scriptsize
%\begin{center}
%\setlength{\belowcaptionskip}{10pt}
\caption{\label{tab2} The predicted abundances for the BBN primordial light elements. The observational data are listed for comparison.}
%\begin{ruledtabular}
%\begin{tabular}{|c|c|c|c|c|c|}
\begin{tabular}{ccccccc}
%\tableline\tableline
\hline \hline
Nuclide & \citet{coc12} & \citet{cyb16} & \citet{ber13} & \multicolumn{2}{c}{This work} & Observation \\
\cline{5-6}
& ($q$=1) & ($q$=1) & ($q$=1) & $q$=1 & $q$=1.069$\sim$1.082 & \\
\hline
$^4$He                       & 0.2476 & 0.2470 & 0.249 & 0.247 & 0.2469         & 0.2561$\pm$0.0108~\citep{ave10} \\
D/H($\times$10$^{-5}$)       & 2.59   & 2.58   & 2.62  & 2.57  & 3.14$\sim$3.25 & 3.02$\pm$0.23~\citep{oli12}     \\
$^3$He/H($\times$10$^{-5}$)  & 1.04   & 1.00   & 0.98  & 1.04  & 1.46$\sim$1.50 & 1.1$\pm$0.2~\citep{ban02}       \\
$^7$Li/H($\times$10$^{-10}$) & 5.24   & 4.65   & 4.39  & 5.23  & 1.62$\sim$1.90 & 1.58$\pm$0.31~\citep{sbo10}     \\
%\tableline\tableline
\hline \hline
\end{tabular}
%\end{ruledtabular}
\end{table*}

\begin{figure}[tbp]
\begin{center}
\includegraphics[width=8.6cm]{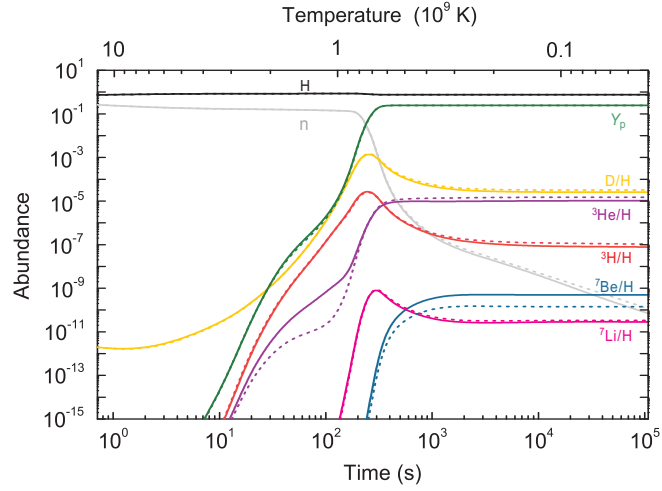}
\vspace{-1mm}
\caption{\label{fig3} Time and temperature evolution of primordial light-element abundances during the BBN era. The solid and dotted lines indicate the results for
the classical MB distribution ($q$ = 1) and the non-extensive distribution ($q$ = 1.0755), respectively.}
\end{center}
\end{figure}

The agreement of our predicted $^7$Li abundance with observations can be attributed to the reduced production of $^7$Li and radioactive $^7$Be (which decays to $^7$Li)
when $q>1$. Production of these species is dominated by the radiative capture reactions $^3$H($\alpha$,$\gamma$)$^7$Li and $^3$He($\alpha$,$\gamma$)$^7$Be, respectively.
The forward alpha-capture rates of these reactions decrease for $q>1$ due to the decreased availability of high energy baryons relative to the MB ($q$ = 1) distribution
(see Fig. 2). On the other hand, the reverse photodisintegration rates are independent of $q$ due to our adoption of Planck's radiation law for the energy density of
photons. As a result, the net production of $^7$Li and $^7$Be decreases, giving rise to concordance between predicted and observed primordial abundances.
Figure~\ref{fig3} shows the time and temperature evolution of the primordial abundances during BBN calculated with the MB and the non-extensive distributions (with
average value of $q$ allowed, $q$ = 1.0755). It can be seen that the predicted $^7$Be (ultimately decaying to $^7$Li) abundance with $q$ = 1.0755 is reduced
significantly relative to that with $q$ = 1, and ultimately the $^7$Li problem can be solved in this model.

\begin{figure}[tbp]
\begin{center}
\includegraphics[width=8.6cm]{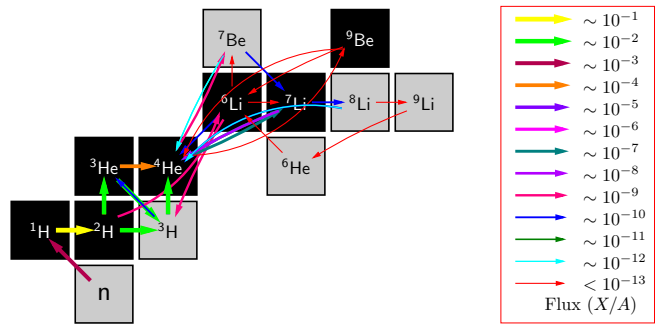}
\vspace{-1mm}
\caption{\label{fig4} The time-integrated fluxes for primary reactions involved in BBN, as calculated using a non-extensive velocity distribution with $q$ = 1.0755.}
\end{center}
\end{figure}

The time-integrated reaction fluxes are calculated within the frameworks of classical MB and non-extensive distributions, respectively. Figure~\ref{fig4} displays the
reaction network for the most important reactions that occur during BBN with a non-extensive parameter of $q$ = 1.0755, where the reaction fluxes are scaled by the
thickness of the solid lines. It demonstrates, in particular, that for $q$ within our allowed range, the fluxes of the main reactions responsible for the net
production of $^7$Be (such as $^3$He($\alpha$,$\gamma$)$^7$Be and $^7$Be($n$,$p$)$^7$Li) are reduced by about 60\% relative to fluxes determined using $q$ = 1.
Thus, it results in an ultimate smaller predicted $^7$Li abundance, which is consistent with observations. The corresponding flux ratios are listed in Table~\ref{tab1}.

One can rationalize the above modified statistics based upon the following arguments. Since the nuclear reactions that lead to the production of $^7$Li and $^7$Be
occur during the end of BBN, they are falling out of equilibrium and must be evolved via the Boltzmann equation. In general, the Boltzmann equations become a coupled
set of partial-integral differential equations for the phase-space distributions and scattering of all species present. Here, we can reduce our consideration to the
evolution of the distribution functions of the $A=3,4$ species contributing to the formation of $A=7$ isotopes. For these species there are two competing processes.
On the one hand the nuclear reaction cross sections favor the reactions among the most energetic $^3$He, $^3$H, and $^4$He nuclei which would tend to diminish slightly
the distributions in the highest energies. At the same time however, the much more frequent scattering of these nuclei off of ambient electrons and (to a lesser
extent) photons will tend to restore the distributions to equilibrium. The competition between these two processes, plus the fact that the distributions of $^3$He,
$^3$H are Fermi-Dirac will lead to a slight deviation from standard MB statistics.

\section{Conclusion}
We have studied the impact on BBN predictions of adopting a generalized distribution to describe the velocities of nucleons in the framework of Tsallis non-extensive
statistics. By introducing a non-extensive parameter $q$, we find excellent agreement between predicted and observed primordial abundances of D, $^4$He and $^7$Li
in the region of $1.069\leq q \leq 1.082$ ($q$ = 1 indicating the classical Maxwell-Boltzmann distribution), which might suggest a possible new solution to the
cosmological lithium problem. We encourage studies to examine sources for departures from classical thermodynamics during the BBN era so as to assess the viability
of this mechanism. Furthermore, the implications of non-extensive statistics in other astrophysical environments should be explored as this may offer new insight
into stellar nucleosynthesis.

\acknowledgments

This work was financially supported by the the National Natural Science Foundation of China (Nos. 11490562, 11675229), and the
Major State Basic Research Development Program of China (2016YFA0400503). AP was partially supported by the Spanish MICINN under Grant No. AYA2013-42762.
CB acknowledges support under U.S. DOE Grant DDE- FG02- 08ER41533, U.S. NSF grant PHY-1415656, and the NSF CUSTIPEN grant DE-FG02-13ER42025.


\begin{thebibliography}{}
\bibitem[Ahmad et al.(2001)]{ahm01} Ahmad, Q.R., et al., 2001, \prl, 87, 071301
\bibitem[Anders et al.(2014)]{and14} Anders, M., et al., 2014, \prl, 113, 042501
\bibitem[Angulo et al.(1999)]{ang99} Angulo, C., et al., 1999, \nphysa, 656, 3
\bibitem[Angulo et al.(2005)]{ang05} Angulo, C., et al., 2005, \apj, 630, L105
\bibitem[Asplund et al.(2006)]{asp06} Asplund, M., et al., 2006, \apj, 644, 229
\bibitem[Aver et al.(2010)]{ave10} Aver, E., et al., 2010, J. Cosmol. Astro-Particle Phys., 5, 003
\bibitem[Bania et al.(2002)]{ban02} Bania, T.M., et al., 2002, Nature, 415, 54
\bibitem[Bertulani et al.(2013)]{ber13} Bertulani, C.A., et al., 2013, \apj, 767, 67
\bibitem[Boyd et al.(2010)]{boy10} Boyd, R.N., et al., 2010, \prd, 82, 105005
\bibitem[Caughlan \& Fowler(1988)]{cf88} Caughlan, G.R. \& Fowler, W.A. 1988, At. Data Nucl. Data Tables, 40, 283
\bibitem[Chakraborty et al.(2011)]{cha11} Chakraborty, N., et al., 2011, \prd, 83, 063006
\bibitem[Clayton et al.(1975)]{cly75} Clayton, D.D., et al., 1975, \apj, 199, 494
\bibitem[Cyburt et al.(2003)]{cyb03} Cyburt, R.H., et al., 2003, Phys. Lett. B, 567, 227
\bibitem[Cyburt et al.(2008)]{cyb08} Cyburt, R.H., et al., 2008, J. Cosmol. Astro-Particle Phys., 11, 012
\bibitem[Cyburt \& Pospelov(2009)]{cyb09} Cyburt, R.H. \& Pospelov, M. 2009, arXiv: 0906.4373
\bibitem[Cyburt et al.(2016)]{cyb16} Cyburt, R.H., et al., 2016, Rev. Mod. Phys., 88, 015004
\bibitem[Coc et al.(2004)]{coc04} Coc, A., et al., 2004, \apj, 600, 544
\bibitem[Coc et al.(2012)]{coc12} Coc, A., et al., 2012, \apj, 744, 158
\bibitem[Coc et al.(2013)]{coc13} Coc, A., et al., 2013, \prd, 87, 123530
\bibitem[Descouvemont et al.(2004)]{des04} Descouvemont, P., et al., 2004, At. Data Nucl. Data Tables, 88, 203
\bibitem[Dubovichenko(2009)]{dub09} Dubovichenko, S.B. 2009, Rus. Phys. J., 52, 294
\bibitem[Famiano et al.(2016)]{fam16} Famiano, M.A., et al., 2016, \prc, 93, 045804
\bibitem[Fields(2011)]{fie11} Fields, B.D. 2011, Ann. Rev. Nucl. Part. Sci., 61, 47
\bibitem[Gamow(1946)]{gam46} Gamow, G. 1946, Phys. Rev., 70, 572
\bibitem[Gell-Mann \& Tsallis(2004)]{gel04} Gell-Mann, M. \& Tsallis, C. 2004, Nonextensive Entropy: Interdisciplinary Applications (Oxford University Press, New York)
\bibitem[Goudelis et al.(2016)]{and16} Goudelis, A., et al., 2016, \prl, 116, 211303
\bibitem[Hammache et al.(2013)]{ham13} Hammache, F., et al., 2013, \prc, 88, 062802(R)
\bibitem[Hara et al.(2003)]{har03} Hara, K.Y., et al., 2003, \prd, 68, 072001
\bibitem[He et al.(2013)]{hjj13} He, J.J., et al., 2013, Phys. Lett. B, 725, 287
\bibitem[Hinshaw et al.(2013)]{WMAP9} Hinshaw, G., et al., 2013, \apjs, 208, 19
\bibitem[Iliadis(2007)]{ili07} Iliadis, C. 2007, Nuclear Physics of Stars (Wiley, Weinheim)
\bibitem[Kang et al.(2012)]{kan12} Kang, M.M., et al., 2012, J. Cosmol. Astro-Particle Phys., 05, 011
\bibitem[King et al.(1977)]{kin77} King, C.H., et al., 1977, \prc, 16, 1712
\bibitem[Kirsebom \& Davids(2011)]{kir11} Kirsebom, O.S. \& Davids, B. 2011, \prc, 84, 058801
\bibitem[Korn et al.(2006)]{kor06} Korn, A.J., et al., 2006, Nature, 442, 657
\bibitem[Kusakabe et al.(2014)]{kus14} Kusakabe, M., et al., 2014, \apjs, 214, 5
\bibitem[Li et al.(2005)]{li05} Li, Z.H., et al., 2005, \prc, 71, 052801(R)
\bibitem[Malaney \& Fowler(1989)]{mal89} Malaney, R.A. \& Fowler, W.A. 1989, \apj, 345, L5
\bibitem[Mathews et al.(2011)]{mat11} Mathews, G.J., et al., 2011, \apj, 727, 10
\bibitem[Miller \& Kusch(1955)]{mil55} Miller, R.C. \& Kusch, P. 1955, Phys. Rev., 99, 1314
\bibitem[Olive et al.(2012)]{oli12} Olive, K.A., et al., 2012, Mon. Not. R. Astron. Soc., 426, 1427
\bibitem[Olive et al.(2014)]{oli14} Olive, K.A., et al., 2014, (Particle Data Group), Chin. Phys. C, 38, 090001
\bibitem[Parker(1972)]{par72} Parker, P.D. 1972, \apj, 175, 261
\bibitem[Penzias \& Wilson(1965)]{pen65} Penzias, A.A. \& Wilson, R.W. 1965, \apj, 142, 419
\bibitem[Pinsonneault et al.(1999)]{pin99} Pinsonneault, M.H., et al., 1999, \apj, 527, 180
\bibitem[Pinsonneault et al.(2002)]{pin02} Pinsonneault, M.H., et al., 2002, \apj, 574, 398
\bibitem[Pizzone et al.(2014)]{piz14} Pizzone, R.G., et al., 2014, \apj, 786, 112
\bibitem[Pospelov \& Pradler(2010)]{pos10} Pospelov, M. \& Pradler, J. 2010, Ann. Rev. Nucl. Part. Sci., 60, 539
\bibitem[Richard et al.(2005)]{ric05} Richard, O., et al., 2005, \apj, 619, 538
\bibitem[Rolfs \& Rodney(1988)]{rol88} Rolfs, C.E. \& Rodney, W.S. 1988, Cauldrons in the Cosmos (Univ. of Chicago Press, Chicago)
\bibitem[Serpico et al.(2004)]{ser04} Serpico, P.D., et al., 2004, J. Cosmol. Astro-Particle Phys., 12, 010
\bibitem[Sbordone et al.(2010)]{sbo10} Sbordone, L., et al., 2010, Astron. Astrophys., 522, A26
\bibitem[Scholl et al.(2011)]{sch11} Scholl, C., et al., 2011, \prc, 84, 014308
\bibitem[Smith et al.(1993)]{smi93} Smith, M.S., et al., 1993, \apjs, 85, 219
\bibitem[Thomas et al.(1993)]{tho93} Thomas, T., et al., 1993, \apj, 406, 569
\bibitem[Torres et al.(1997)]{tor97} Torres, D.F., et al., 1997, \prl, 79, 1588
\bibitem[Torres et al.(1998)]{tor98} Torres, D.F., et al., 1998, \prl, 80, 3889
\bibitem[Tsallis(1988)]{tsa88} Tsallis, C. 1988, J. Stat. Phys., 52, 479
\bibitem[Vauclair \& Charbonnel(1998)]{vau98} Vauclair, S. \& Charbonnel, C. 1998, \apj, 502, 372
\bibitem[Voronchev et al.(2012)]{tor12} Voronchev, V.T., et al., 2012, \prd, 85, 067301
\bibitem[Wagoner(1969)]{wag69} Wagoner, R.V. 1969, \apjs, 18, 247
\bibitem[Wang et al.(2011)]{wan11} Wang, B., et al., 2011, \prc, 83, 018801
\bibitem[Xu et al.(2013)]{xu13} Xu, Y., et al., 2013, \nphysa, 918, 61
\bibitem[Yamazaki et al.(2014)]{yam14} Yamazaki, D.G., et al., 2014, \prd, 90, 023001
%\bibitem[Tsallis et al.(1998)]{tsa98} Tsallis, C., et al., 1998, Physica A, 261, 534
%\bibitem[Livadiotis \& McComas]{liv09} Livadiotis, G. \& McComas, D.J. 2009, J. Geophys. Res., 114, A11
%\bibitem[O'Meara et al.(2006)]{mea06} O'Meara, J.M., et al., 2006, \apj, 649, L61
\end{thebibliography}
\end{document}